\documentstyle[prl,aps,epsf]{revtex}

\begin{document}

\title{Cascades in helical turbulence}

\author{
P. D. Ditlevsen$^1$ and P. Giuliani$^2$\\
$^1$The Niels Bohr Institute, Department for Geophysics,
University of Copenhagen, Juliane Maries Vej 30,
 DK-2100 Copenhagen O, Denmark.\\
$^2$Dipartimento di Fisica and Instituto Nazionale di Fisica della Materia, 
Universit\`{a} della Calabria, 87036 Rende (CS), 
Italy}
\date{\today}
\maketitle
\begin{abstract}
The existence of a second quadratic inviscid invariant, the helicity, in
a turbulent flow leads to 
coexisting cascades of energy and helicity.
An equivalent of the four-fifth law for the longitudinal third order
structure function, which is derived from energy conservation, 
is easily derived from helicity conservation \cite{Procaccia,russian}. 
The ratio of dissipation of helicity to dissipation of energy is proportional
to the wave-number leading to a different Kolmogorov scale for helicity than
for energy. 
The Kolmogorov scale for helicity is always larger
than the Kolmogorov scale for energy so in the high Reynolds number limit the
flow will always be helicity free in the small scales, much in the same
way as the flow will be isotropic and homogeneous in the small scales.
A consequence is that a pure helicity
cascade is not possible. The idea is illustrated in a shell model of
turbulence.
\end{abstract}

Few exact results regarding fully developed turbulence have yet
been derived.  The most celebrated being Kolmogorovs four-fifth law
\cite{Frisch}. The four-fifth law is based on the fact that energy, which
is an inviscid invariant of the flow, is transferred through the inertial
range from the integral scale to the dissipation scale.  The four-fifth
law, $\langle \delta v(l)_\|^3\rangle = -(4/5) \overline{\varepsilon}l$, states
that the third order correlator associated with energy flux equals the
mean energy dissipation.  As noted recently \cite{Procaccia,russian} in
the case of helical flow a similar relation exists for the transfer of
helicity leading to an other scaling relation for a third order correlator
associated with the flux of helicity, 
$\langle \delta {\bf v}_\|(l)\cdot[{\bf v}_\bot(r)\times {\bf v}_\bot(r+l)]\rangle 
= (2/15) \overline{\delta}l^2$, where $\overline{\delta}$ is the mean dissipation
of helicity. This relation is called the 'two-fifteenth
law' due to the numerical prefactor. 
This establishes another non-trivial
scaling relation for velocity differences in a turbulent helical flow. 

The coexistence of cascades of energy and enstrophy is prohibited
for high Reynolds number flow in 2D turbulence. 
The reason for this is that the enstrophy dominates at small scales such
that the ratio of energy -- to enstrophy dissipation vanishes for high
Reynolds number flow. The Kolmogorov scale $k_Z^{-1}$ for enstrophy dissipation is
determined from the energy spectrum $E(k)\sim k^{-3}$ and the kinematic
viscosity $\nu$ by $\overline{\zeta}=\nu \int^{k_Z}dkk^4E(k)\sim \nu k_Z^2 \Rightarrow
k_Z \sim \nu^{-1/2}$. The energy dissipation is $\overline{\varepsilon}=
\nu \int^{k_Z}dkk^2E(k)\sim \nu \log k_Z \sim -(1/2) \nu \log \nu \rightarrow
0$ for $\nu\rightarrow 0$. Consequently energy is cascaded upscale in
2D turbulence.

The situation in
3D turbulence is different. Here
coexisting cascades of energy and helicity are possible \cite{Lesieur}. 
However, the same type of dimensional argument as for the
cascades of energy and enstrophy in 2D turbulence applies. The helicity
density is $h=u_i\omega_i$, where $\omega_i=\epsilon_{ijk}\partial_ju_k$ is
the vorticity. The mean dissipation of helicity is 
$D_H=\nu \langle \partial_j u_i \partial_j \omega_i\rangle$.
Disregarding signs this can spectrally be represented as

\begin{equation}
D_H \sim \nu \int^{k_E}dkk^3E(k)\sim \nu k_E^{7/3}\sim
\nu^{-3/4}, 
\label{DH}
\end{equation}
where $k_E^{-1}=\eta$ is the Kolmogorov scale and we have used $E(k)\sim k^{-5/3}$ and $k_E\sim \nu^{-3/4}$. 
This means that for high Reynolds numbers flow the dissipation of helicity will
grow as $Re^{3/4}$. Since the mean dissipations of energy $\overline{\varepsilon}$ and
helicity $\overline{\delta}$ are determined by the integral scale forcing 
the growth of helicity dissipation with Reynolds number is apparently in conflict with the assumption of a constant energy dissipation
in the limit of vanishing viscosity. This is not a true problem because
helicity is non-positive, and the viscous term in the equation for the helicity 
$\nu (u_i \partial_{jj} \omega_i+
\omega_i \partial_{jj} u_i)$ can have either sign. So in the high
Reynolds number limit 
there must either be a detailed balance between dissipation
of positive and negative helicity or the energy cascade is blocked \cite{Levich}. In the rather artificial case of a shell model
where only one sign of helicity is dissipated by hyper-viscosity, the energy cascade
is indeed prevented all together similar to the case of forward energy cascade of energy in 2D turbulence \cite{PD}.

In a helical flow $(\overline{\delta}\ne 0)$
the dissipation of helicity defines a scale different from the Kolmogorov scale $\eta$.
This we will call the Kolmogorov scale $\xi$ for 
dissipation of helicity.

Following K41, the Kolmogorov scale $\eta$ for energy dissipation 
is obtained from $\overline{\varepsilon}
\sim \delta u_\eta^3/\eta\sim \nu \delta u_\eta^2/\eta^2
\Rightarrow
\eta\sim (\nu^3/\overline{\varepsilon})^{1/4}$, where $\delta u_l$ is a
typical variation of the velocity over a scale $l$.
The Kolmogorov scale $\xi$ for dissipation of helicity is defined as the scale 
where the helicity dissipation 
is of same order as the
spectral helicity flux.  With dimensional counting 
we have $\overline{\delta}\sim \nu \delta u_\xi^3/\xi^2$ and
using $\delta u_l \sim (l \overline{\varepsilon})^{1/3}$ we obtain

\begin{equation}
\xi \sim (\nu^3 \overline{\varepsilon}^2/\overline{\delta}^3)^{1/7}.
\label{KH}
\end{equation}

Now it is clear why (\ref{DH}) leads to a wrong conclusion for the mean dissipation
of the helicity $\overline{\delta}$. 
The integral will not be dominated by contributions from $k_E$ but
contributions from $k_H=1/\xi$, 

\begin{equation}
D_H= \overline{\delta} \sim \nu k_H^{7/3} \Rightarrow k_H\sim \nu^{-3/7}.
\end{equation}

The ratio of the two Kolmogorov scales is then $(\eta/\xi)=(k_H/k_E)\sim \nu^{-3/7+3/4}=\nu^{9/28}\rightarrow 0$
for $\nu\rightarrow 0$. Thus for high Reynolds number helical flow the small scales will always
be non-helical and a pure helicity cascade is not possible.

On the other hand for scales $l < \xi$ the ratio of dissipation of energy and helicity is
proportional to $l$, $D_E/D_H \sim l$, which means that helicity dissipation dominates and
the dissipation of positive
and negative helicity must balance.
The reason for the flow to be non-helical on small scales is different from the reason
why the flow tends to be isotropic on small scales even though the integral scale is
non-isotropic. The reason for the small scales to be isotropic is that the structure functions
associated with the non-isotropic sectors scale with scaling exponents that are larger than
those of the isotropic sector and thus becomes sub-leading for the flow at small scales
independent of the dissipation \cite{Arad}.

The physical 
picture for fully developed helical turbulence is then
that $\overline{\delta}$ and $\overline{\varepsilon}$ are solely determined by
the forcing in the integral scale. There will then be an inertial range with
coexisting cascades of energy and helicity with third order structure functions
determined by the four-fifth -- and the two-fifteenth laws. This is followed
by an inertial range between $\xi$ and $\eta$ corresponding to non-helical turbulence,
where the dissipation of positive and negative helicity vortices balance and
the two-fifteenth law is not applicable.

In order to test these ideas in a model system we investigate the role
of helicity and the structure of the helicity transfer
in a shell model.

Shell models are
toy-models of turbulence which by construction have second order inviscid
invariants similar to energy and helicity in 3D turbulence. 
Shell models
can be investigated numerically for high Reynolds numbers,
in contrast to the Navier-Stokes equation, and high order statistics
and anomalous scaling exponents are easily accessible. 
Shell models lack any spatial structures so
we stress that only certain aspects of the turbulent cascades have
meaningful analogies in the shell models. This should especially
be kept in mind when studying helicity which is intimately linked
to spatial structures, and the dissipation of helicity to reconnection of
vortex tubes \cite{Levich}. So the following only concerns the spectral aspects
of the helicity and energy cascades.

The most well studied shell model,
the GOY model \cite{GOY}, is defined from the governing equation,

\begin{equation}
\dot{u_n}=i k_n (u_{n+2}u_{n+1}-\frac{\epsilon}{\lambda}u_{n+1}u_{n-1}+
\frac{\epsilon-1}{\lambda^2}u_{n-1}u_{n-2})^* -\nu k_n^2 u_n + f_n
\label{1}
\end{equation}
with $n=1, ..., N$ where the $u_n$'s are the complex shell velocities. The
wavenumbers are defined as $k_n = \lambda^n$, where $\lambda$ is the shell
spacing. The second and third terms are dissipation and forcing. The model
has two inviscid invariants, energy, $E=\sum_n E_n =\sum_n |u_n|^2$,
and 'helicity', $H=\sum_nH_n=\sum_n (\epsilon -1)^{-n}|u_n|^2$. The
model has two free parameters, $\lambda$ and $\epsilon$. The 'helicity'
only has the correct dimension of helicity if $|\epsilon -1|^{-n}=k_n
\Rightarrow 1/(1-\epsilon)=\lambda$. 
In
this work we use the standard parameters $(\epsilon,\lambda)=(1/2,2)$
for the GOY model.

A natural way to define the structure functions of moment $p$ is through the 
transfer rates of the inviscid invariants,

\begin{eqnarray}
S^E_p(k_n)= \langle(\Pi^E_n)^{p/3}\rangle k_n^{-p/3} \\
S^H_p(k_n)= \langle(\Pi^H_n)^{p/3}\rangle k_n^{-2p/3}
\label{s3h}
\end{eqnarray}   
The energy flux is defined in the usual
way as $\Pi^E_n = d/dt|_{n.l.}(\sum_{m=1}^{n} E_m) $ where
$d/dt|_{n.l.}$ is the time rate of change due to the non-linear term in
(\ref{1}). The helicity flux $\Pi^H_n$ is defined similarly. By a simple algebra we have
the following expression for the fluxes,

\begin{eqnarray}
\langle \Pi^E_n \rangle= (1-\epsilon) \Delta_n + \Delta_{n+1} =\overline{\varepsilon}\label{pie} \\
\langle \Pi^H_n \rangle= (-1)^n k_n(\Delta_{n+1}-\Delta_n)=\overline{\delta}
\label{pih}
\end{eqnarray}
where $\Delta_n = k_{n-1}Im \langle u_{n-1}u_nu_{n+1}\rangle$,
$\overline{\varepsilon}$ and $\overline{\delta}$ are the mean dissipations of
energy and helicity respectively. The first equalities hold without averaging
as well.
These equations are the shell model equivalents of
the four-fifth -- and the two-fifteenth law.
Kadanoff et al. \cite{Kadanoff}
defined a third order structure function as, 

\begin{equation}S^3_n = Im\langle u_{n-1}u_nu_{n+1}\rangle =\Delta_n/k_{n-1}
\label{s3n1}\end{equation}
to avoid the spurious (specific to the GOY model) period 3 oscillation. Using
this we obtain from (\ref{pie}) and (\ref{pih}) a scaling relation for $S_n^3$,

\begin{equation}
S^3_n = \frac{1}{(1-\epsilon/2)k_n}(\overline{\varepsilon}-(-1)^n \overline{\delta}/k_n).
\label{s3n2}
\end{equation}
The last term in the parenthesis is sub-leading with period two oscillations.
When $\overline{\delta}=0$ the sub-leading term disappears and the scaling 
from the four-fifth law is obtained, figure 1. The relation (\ref{pih}) is
the scaling relation corresponding to the sub-leading term, which survives
due to detailed cancellations between the two terms $\Delta_{n+1}$ and 
$\Delta_n$ of the leading term corresponding to (\ref{pie}). 
The case $\overline{\varepsilon}=0$ and
$\overline{\delta} \ne 0$ would, aside from the period two oscillation, correspond to a helicity cascade with the
scaling obtained from dimensional counting $u_n \sim k_n^{-2/3}$. However, this
situation is, as we will show shortly, not realizable.   

The mean dissipations $\overline{\varepsilon}$ and $\overline{\delta}$ are from 
energy and
helicity conservations identical to the mean energy and helicity
inputs which from (\ref{1}) are, 
\begin{equation}
\overline{\varepsilon}=\sum_n
\langle f_n u_n^*\rangle + c.c.
\label{eforce}
\end{equation}
and 

\begin{equation}
\overline{\delta}=\sum_n
(-1)^nk_n \langle f_n u_n^*\rangle + c.c.,
\label{hforce}
\end{equation}
so $\overline{\varepsilon}$
and $\overline{\delta}$ are not independent. The forcing can be chosen
in many ways. A natural choice is
$f_n= f_n^0/u_n^*$, where $f_n^0$ is independent on the shell velocities.
Then we have, $\overline{\varepsilon}=\sum_{n<n_{I}}f_n^0$ and
$\overline{\delta}=\sum_{n<n_{I}}(-1)^nk_n f_n^0$, $n_{I}$
indicates the end of the integral scale. By choosing
the coefficients, stochastic or deterministic functions of time, this last sum can 
vanish identically, which is referred to as helicity free
forcing. The simulations shown in figure 1 are performed with the forcing,
$f_3^0= 10^{-2}(1+i)$ and $f_4^0 = -A f_3^0/\lambda$ with $A=0$ and
$A=1$, corresponding to $(\overline{\varepsilon},\overline{\delta})=(0.01,0)$
and $(\overline{\varepsilon},\overline{\delta})=(0.01,0.08)$ respectively. 

Helicity is not positive and is dissipated with opposite signs
for odd and even shells. 
If we consider the third order structure function 
associated with the helicity transfer as defined by (\ref{s3h})
we see (figure 2) period two oscillations growing with $n$.  This
period two oscillation is due to the dissipation and not the non-linear
transfer. 
The helicity flux is

\begin{equation}
\langle \Pi^H_n \rangle =  \overline{\delta} - \langle D_n \rangle,
\end{equation}
where $D_n$ is the helicity dissipation at shells $m\le n$:

\begin{equation}
D_n = \sum_{m=1}^n \nu (-1)^m k_m^3 |u_m|^2.
\label{Dh}
\end{equation}
In the inertial range for energy transfer we have the Kolmogorov
scaling $u_n\sim k_n^{-1/3}$ so the helicity dissipation can be estimated,

\begin{equation}
D_n \sim \sum_{m=1}^n \nu (-1)^m k_m^{7/3} \sim \lambda^{7/3} \frac{
(-1)^n \lambda^{7n/3}-1}{\lambda^{7/3}+1} \sim (-1)^nk_n^{7/3}.
\label{diss}
\end{equation}
Figure 3 shows $|\langle\Pi^H_n\rangle|$ and $\langle\Pi^E_n\rangle$ as functions of wave number.
The scaling (\ref{diss}) of the helicity dissipation is the straight line, 
the horizontal dashed line is ${\overline{\delta}}$. The inertial range for
helicity transfer is to the left of the crossing of the two lines.
The crossing is the Kolmogorov scale for helicity transfer $K_H$, which
does not coincide with the Kolmogorov scale for energy transfer, $K_E$.
The 'pile-up' for $k$ larger than $K_H$ was earlier interpreted as a bottleneck effect
\cite{Biferale}. It is a
balance between positive and negative helicity dissipation.
The forcing $f_n=f_n^0/u_n^*$ can potentially cause numerical trouble
when $|u_n|$ becomes small. It is easy to see that the linear
equation for (real) shell velocity $u_n$, neglecting the non-linear transfer, $\dot{u}_n=
f/u_n$ will create a finite time singularity. This is not the case for the 
forcing suggested by Olla \cite{Olla} at two shells, $f_n=(a E_{n+1}u_n)/(E_{n}+E_{n+1})$
and $f_{n+1}=(b E_nu_{n+1})/(E_n+E_{n+1})$, where $a$ and $b$ are constants determining the
ratio of energy -- to helicity input. The coupled set of equations,
$(\dot{u}_n=f_n,\, \dot{u}_{n+1}=f_{n+1})$ is integrable (solve for $y=u_n/u_{n+1}$), and
has no finite time singularities. Using this forcing we performed a set of simulations
with constant energy input $\overline{\varepsilon}=0.01$ and varying helicity input
$\overline{\delta}=(0.0001,0.001,0.005,0.01,0.08)$. In figure 4 the spectra of the
absolute value of the helicity transfer normalized with $\overline{\delta}$ are plotted against wave number
normalized with $K_H$. $K_H$ is in each case calculated from (\ref{KH}), and a clear
data collapse is seen. 
 
In summary the simulations with the GOY shell model suggest a new
Kolmogorov scale for helicity, always smaller than the Kolmogorov
scale for energy. Thus there exist two inertial ranges in 
helical turbulence, a range smaller than $K_H$ with coexisting
cascades of energy and helicity where both the four-fifth - and the two-fifteenth
law applies, and a range between $K_H$ and $K_E$ where the flow is 
non-helical and only the four-fifth law applies.

\newpage
\begin{center}
FIGURE CAPTIONS
\end{center}
\newcounter{fig}
\begin{list}{Fig. \arabic{fig}}
{\usecounter{fig}\setlength{\labelwidth}{2cm}\setlength{\labelsep}{3mm}}

\item The third order structure function $S^3_n$ as calculated from 
(\ref{s3n1}) in the cases $\overline{\delta}>0$ (crosses) and $\overline{\delta}=0$ 
(diamonds).
In the case of helicity free forcing the modulus 2
oscillations disappears. 
In the two runs we have 25 shells, $\nu=10^{-9}, f_n=0.01 (1+i)(\delta_{n,2}/u_2^*
-A\delta_{n,3}/2u_3^*)$ with $A=0,1$ respectively.

\item The helicity flux $\langle\Pi^H_n\rangle$ in the case $\overline{\delta}>0$.
The same curve
is multiplied by 1000
and over-plotted in order to see the inertial range. 
The period 2 oscillations in the helicity transfer comes from the
helicity dissipation.

\item The absolute values of the helicity flux $|\langle\Pi^H_n\rangle|$ (diamonds)
show
a crossover from the inertial range for helicity to the range where the
helicity is dissipated. The line has a slope of $7/3$ indicating
the helicity dissipation.  The dashed lines indicate the helicity
input $\overline{\delta}$. The crosses is the helicity flux in the case
$\overline{\delta}=0$ where there is no inertial range and $K_H$ coincides
with the integral scale.
The triangles are the energy flux $\langle\Pi^E_n\rangle$.

\item Five simulations with constant viscosity $\nu=10^{-9}$, constant 
energy input $\overline{\varepsilon}=0.01$
and varying helicity input $\overline{\delta}=(0.0001,0.001,0.005,0.01,0.08)$ are shown. 
The
absolut values of the helicity flux $|\langle\Pi^H_n\rangle|$ divided by
$\overline{\delta}$ is plotted against the wave number divided by 
$K_H=(\nu^3\overline{\varepsilon}^2/\overline{\delta}^3)^{-1/7}$,
which is obtained from (\ref{KH}) neglecting $O(1)$ constants. A clear data
collapse is seen.

\end{list}

\newpage
\begin{figure}[htb]
\epsfxsize=8.5cm
\epsffile{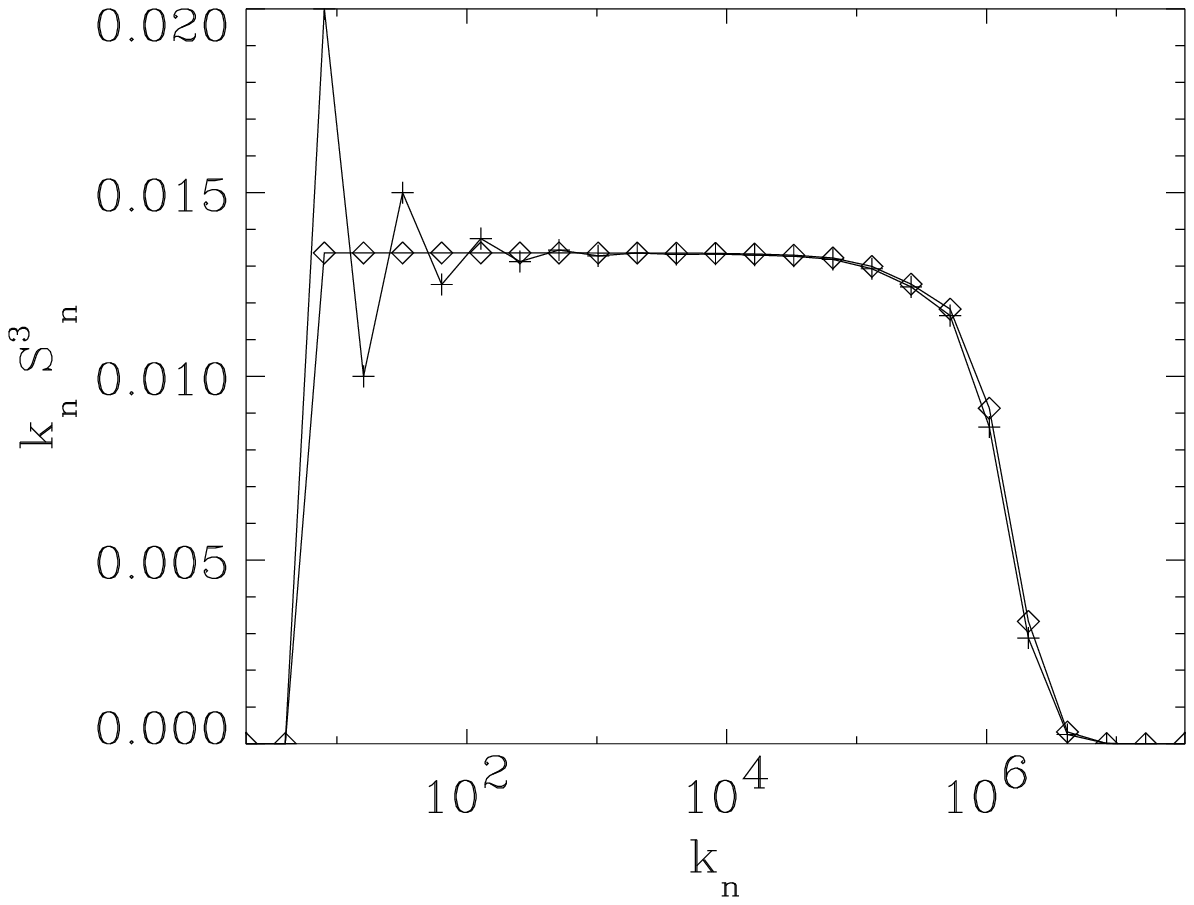}
\caption[]{}
\end{figure}
\begin{figure}[htb]
\epsfxsize=8.5cm
\epsffile{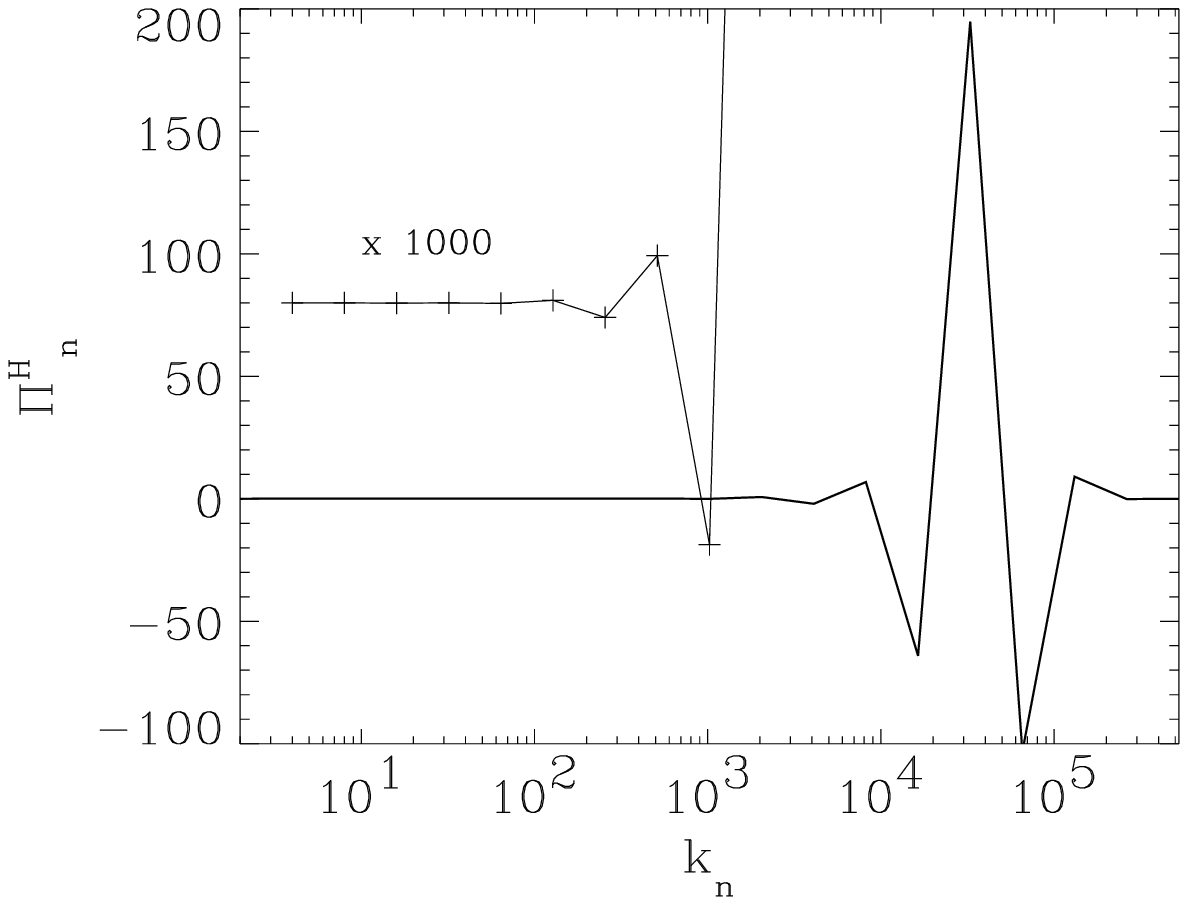}
\caption[]{}
\end{figure}
\begin{figure}[htb]
\epsfxsize=8.5cm
\epsffile{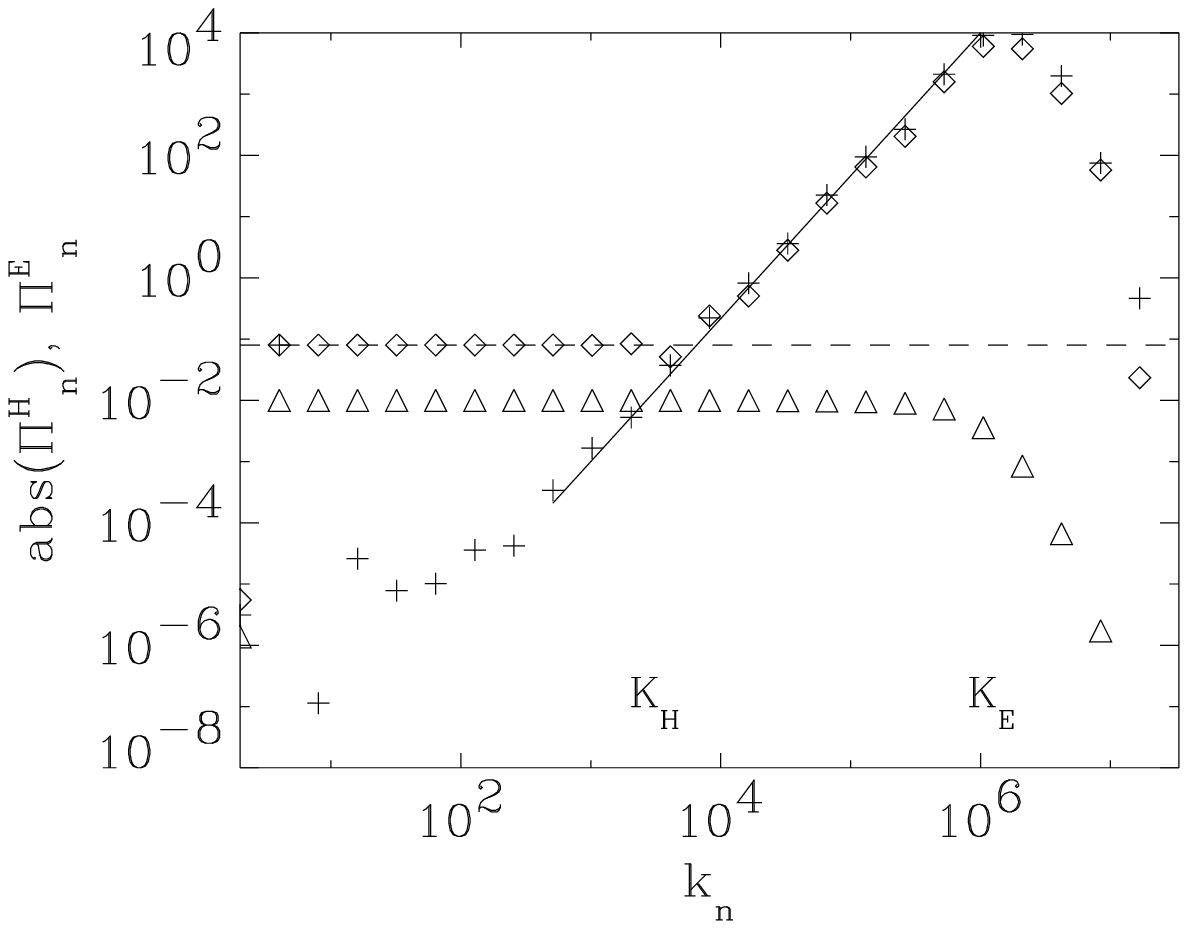}
\caption[]{}
\end{figure}
\newpage
\begin{figure}[htb]
\epsfxsize=8.5cm
\epsffile{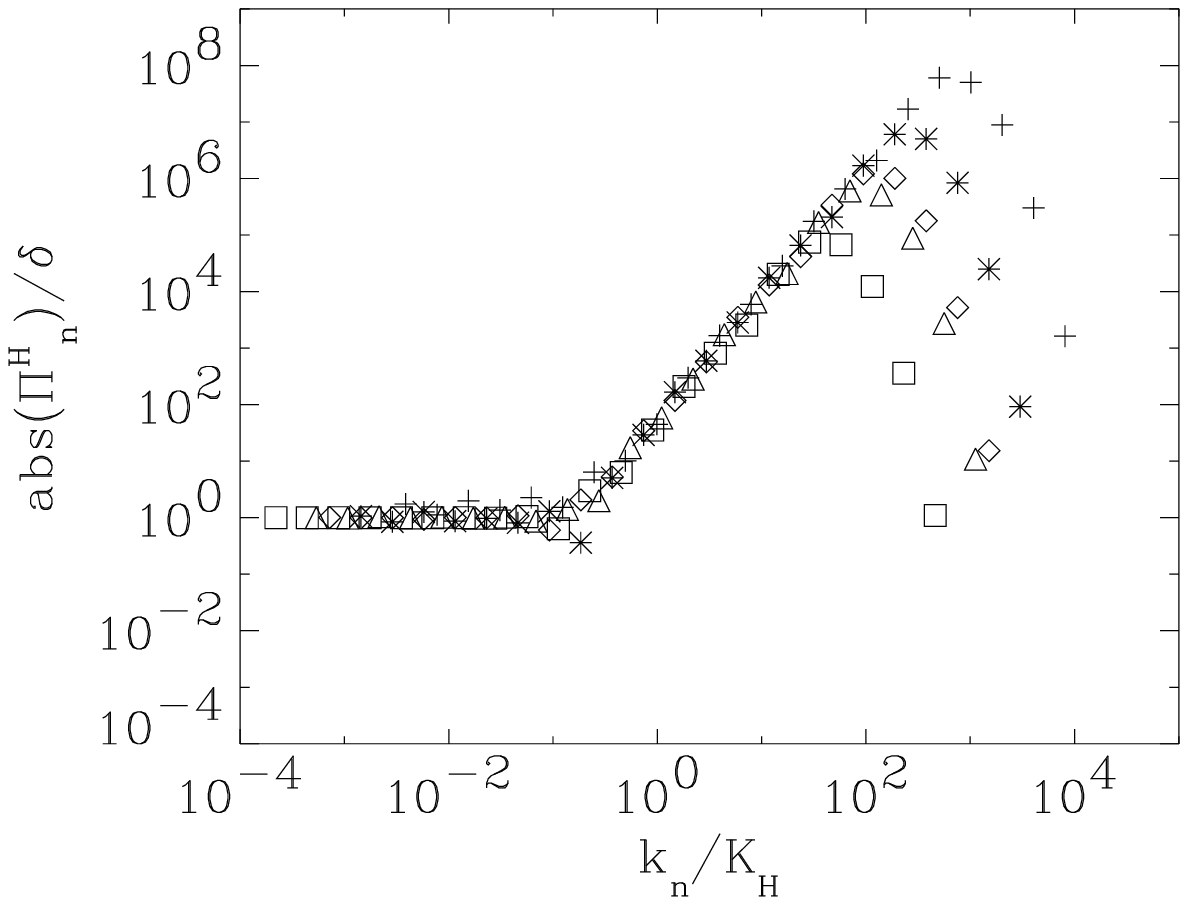}
\caption[]{}
\end{figure}

\end{document}